\documentclass[useAMS,usenatbib]{mn2e}
\usepackage {graphicx}

\title[Could the Galactic disk heating be due to Globular Cluster impacts?]{Could the Galactic disk heating be due to Globular Cluster impacts?}

\author[D. Vande Putte, Mark Cropper and Ignacio Ferreras]{D. Vande Putte,\thanks{E-mail:
dwvp@mssl.ucl.ac.uk (DVP)} Mark Cropper \thanks{E-mail:msc@mssl.ucl.ac.uk (MC)}  and Ignacio Ferreras\thanks{E-mail:
ipf@mssl.ucl.ac.uk (IF)}
\\
University College London, Mullard Space Science Laboratory, Holmbury St Mary, Dorking, RH5 6NT, UK\\}

\begin{document}

\date{Accepted yyyy mm dd. Received  yyyy mm dd; in original form  yyyy mm dd}

\pagerange{\pageref{firstpage}--\pageref{lastpage}} \pubyear{2002}

\maketitle

\label{firstpage}

\begin{abstract}

So far, six mechanisms have been proposed to account for the Galactic disk heating.  Of these, the most important appear to be a combination of scattering of stars by molecular clouds and by spiral arms.  We study a further mechanism, namely, the repeated disk impact of the original Galactic Globular Cluster population up to the present.  We find that Globular Clusters could have contributed at most a small fraction of the current vertical energy of the disk, as they could heat the whole disk to  {$\sigma $}$_{z}$ = $5.5$kms$^{{\rm -} {\rm 1}}$ (c.f. the observed 18 and 39 kms$^{{\rm -} {\rm 1}}$ for the thick and thin disks respectively). We find that the rate of rise of disk heat ($\alpha$=0.22 in  \textit{$\sigma $}$_{z}$  $\sim \,t^{\alpha}$ with \textit t being time), is close to that found for scattering by molecular clouds. 

\end{abstract}


\section{Introduction}

The disk stars in the Galaxy can be described by a DF represented by a Schwarzschild distribution where the number of 
stars with velocity $v$ within a small range $d^{3}v$ is:

\begin{equation}
\label{equa}
f\left( {v} \right)\,d^{3}v\, = \,{\frac{{n_{0} \,d^{3}v}}{{\left( {2\pi}  
\right)^{3 / 2}\sigma _{R} \,\sigma _{\phi}  \,\sigma _{z}}} }\,\exp {\left[ 
{ - \left( {{\frac{{v^{2}_{R}}} {{2\sigma _{R} ^{2}}}} + {\frac{{v^{2}_{\phi 
}}} {{2\sigma _{\phi}  ^{2}}}} + {\frac{{v^{2}_{z}}} {{2\sigma _{z} ^{2}}}}} 
\right)} \right]}
\end{equation}

\noindent
where $n_{0} $ is the number of stars per unit volume, and the $\sigma _{R} 
\,,\,\sigma _{\phi}  \,,\,\sigma _{z} $ are the dispersions in the three 
cylindrical coordinates (Binney \& Tremaine 2008). This ellipsoid's shape is 
similar for all stars in the disk, but its size depends on the stellar 
population. For example, the observed dispersion is three times larger for cool, red 
stars than for hot blue stars. This suggests that stars are 
born on nearly circular, cold, orbits that heat up as time goes by.  The value determined for the current vertical
heating of the thin disk is $18$kms$^{{\rm -} {\rm 1}}$, that for the thick disk is $39$kms$^{{\rm -} {\rm 1}}$ (Binney \& Merrifield, 1998).

The nature of the heating agent is discussed in detail in Binney {\&} 
Tremaine (2008), who list the following possible candidates: encounters with MACHOs from the dark halo, velocity kicks from passing stars, scattering by spiral arms, scattering by molecular clouds, mergers with satellite galaxies, and finally, substructure in the dark halo.  They conclude that the most likely mechanism is a combination of spiral 
transients and scattering by molecular clouds. 

In this paper we examine whether the repeated impact of Globular Clusters (GC) since 
Galaxy formation could be a significant mechanism for disk heating.  In Section 2 we examine the physical processes involved.  Section 3 presents numerical results for the current GCs.  Section 4 shows the results for the entire GC population since Galaxy formation, then Section 5 discusses these results.

\section{Physical processes}

The Galaxy formed $\sim 10^{10}$ years ago, with its population of Globular 
Clusters moving along orbits in the Galactic potential. Assuming 
that the initial disk of the Galaxy was cold, the GC impacts through the 
disk will gradually heat up the disk as a result of dynamical friction. This 
is a process where a heavier object of mass $M$ gravitationally interacts with 
a large number of randomly distributed less massive objects of mass $m$, and 
transfers some of its kinetic energy to the less massive particles. The 
expression for the deceleration was first given by Chandrasekhar (1943), and re-expressed in a convenient form by Binney \& Tremaine (2008) as:

\begin{equation}
\label{equa}
{\frac{{d\overrightarrow {v_{M}}} } {{dt}}}\, = \, - \,{\frac{{4\pi \ln 
\Lambda G^{2}\rho M}}{{v^{3}_{M} }}}{\left[ {\rm{erf}\left( {X} \right)\, - 
\,{\frac{{2X}}{{\sqrt {\pi}} } }e^{ - X^{2}}} \right]}\,\,\,\overrightarrow 
{v_{M}}  
\end{equation}

\noindent
where$\,\overrightarrow {v_{M}}  $ is the velocity of a GC of mass M, the 
individual disk stars having a mass$\,m\ll M$, $\ln \Lambda$ is the Coulomb logarithm, 
whose value is set to 18.1, the value in the Solar neighbourhood (Binney \& Tremaine, 2008). The parameter $X\, \equiv \,v_{M} / \left( {\sqrt {2} \,\sigma}  
\right)$, and $\sigma $ is the velocity dispersion of disk stars. The local mass density is 
\textit{$\rho $}. Equation 2 assumes that the heavier object is a point mass, but 
tests have shown that it is nevertheless a remarkably good description of 
the drag experienced by a finite body orbiting in a stellar system (White, 1978).  As a result, the GC experiences a force directed opposite to the direction of motion, along a 
path of length \textit{$\lambda $} of disk material equal to the GC pathlength through the 
disk. The energy transfer $E_{T}$ is then:

\begin{equation}
\label{equa}
E_{T} \, = \,{\frac{{4\pi \ln \Lambda G^{2}\rho M^{2}}}{{v_{M}^{2} 
}}}\,{\left[ {\rm{erf}\left( {X} \right)\, - \,{\frac{{2X}}{{\sqrt {\pi}} } }e^{ 
- X^{2}}} \right]}\,\lambda 
\end{equation}so that the GC is slowed down and moves to a lower orbit. 
Ultimately, the orbit will decay to the point where the GC dissolves into 
the bulge or disk.  Formally, Equations 2 and 3 are obtained for an infinite uniform medium, so represent an approximation for a GC crossing the Galactic disk.

The dynamical friction experienced by a GC in the halo is small compared to that during its transit through the disk.  For example, in the case of $\omega $Cen, the dynamical friction from a disk transit is two orders of magnitude greater than that it experiences during the halo part of its orbit.

With time, several phenomena occur, which strip the GC of mass: \textit{stellar evolution} which drives mass out of the GC via winds and 
supernovae, \textit{two-body relaxation} where the cumulative effect of gravitational 
interactions between stars occasionally provides sufficient energy for a star 
to escape the cluster, and  \textit{tidal shocks}, where as the GC transits through the disk, it 
experiences a rapidly changing tidal force which can increase the stars' 
kinetic energy to the point that they escape the cluster.  The result of these phenomena on the Galactic GC population has been studied 
extensively by Fall \& Zhang (2001), and Prieto \& Gnedin (2008). The latter 
study the dissolution of GCs in a constant combined halo-disk potential for 
GC masses in excess of 10$^{5}$$ \textit{M}$$_\odot$ without orbital decay. The 
former work back from today's Galactic GC population, to predict the GC-IMF, based 
on a static spherical potential, and accounting for GC mass loss.

\section{Heating due to the existing globular clusters}

There are $\sim$160 GCs in the Galaxy today, with most listed in the Harris 
catalogue (Harris, 1996), of which the latest on-line version is 
available at www.physics.mcmaster.ca/Globular.html, dated 2003. In this section we describe how we calculate the disk heating were it caused exclusively by these 160 GCs.  To do such a calculation requires a knowledge of the GC trajectories, so that energy transfers at disk crossings may be evaluated.  Determining trajectories  in turn requires a knowledge of GC positions and proper motions.  Of the 160 GCs, only a subset of fifty-four have reliably determined proper motions (Moreno \& Pichardo 2006 and 2008, and Kalirai 2007).   We start by determining the heating due to this subset of fifty-four GCs.  It is possible to determine their past trajectories up to the 
formation of the Galaxy  $\sim10^{10} $years ago if we assume that the Galactic potential has remained constant throughout the time since Galaxy formation.  Note that the disk potential is assumed constant, with the dissolving GCs having no influence on disk mass.  We can then establish the characteristics of their crossings over that time 
(position in disk, velocity, time, angle of incidence on disk).

We follow for the orbit calculations the method described in Vande Putte \& Cropper (2009), using the potential proposed by Fellhauer (2006).  This potential has three components, representing a bulge, halo, and disk:

\begin{equation}
\label{equa}
\Phi \, = \,\Phi _{b} \, + \,\Phi _{h} \, + \,\Phi _{d} 
\end{equation}

The contribution of the bulge is given by a Hernquist (1990) potential:

\begin{equation}
\label{eq5}
\Phi _{b} (r)\, = \, - \,{\frac{{GM_{b}} }{{r\, + \,a}}}
\end{equation}

\noindent
where $M_{b} \, = \,3.4\,\times 10^{10}\,M_{ \odot}  $, and $a\, = \,0.7$ 
kpc. 
\noindent

The axisymmetric disk potential is represented by a Miyamoto-Nagai 
(1975) expression:

\begin{equation}
\label{eq3}
\Phi _{d} (R,z)\, = \, - \,{\frac{{GM_{d}} }{{{\left\{ {R^{2}\, + \,{\left[ 
{a\, + \,(z^{2} + b^{2})^{1 / 2}} \right]}^{2}} \right\}}^{1 / 2}}}}
\end{equation}

\noindent 
with $M_{d} \, = \,10^{11}\,\,M_{ 
\odot}  $, $a = \,6.5$kpc, $b\, = \,0.26$kpc. A spherically-symmetric halo contributes to the potential in a logarithmic form:

\begin{equation}
\label{eq6}
\Phi _{h} \, = \,{\frac{{v_{0}^{2}} }{{2}}}\,\ln (R^{2}\, + \,z^{2}\, + \,d^{2})
\end{equation}
\noindent
with $v_{0} \, = 186$kms$^{{\rm -} {\rm 1}}$, 
 and $d\, = \,12$kpc  (Fellhauer et al, 2006).

In our orbit calculations we track the GCs, and note the crossings data such as time, position of crossing, and angle of incidence on the disk.  In doing this, we
found that the fifty-four GCs cross the disk plane 10454 times over the last 10 Gyr. We assume an initial disk star
velocity dispersion in the $z$ direction, of $1$kms$^{{\rm -} {\rm 1}}$, based on the study of vertical velocity dispersion in disks as a function of time, by Jenkins \& Binney (1990), thus finding the initial vertical energy in the disk.
We then use Equation 3 to calculate the energy transfer to the 
disk at each crossing. Vande Putte \& Cropper (2008) show how the track of the GC through the disk focusses stars towards it, essentially randomising the velocities, so we divide this energy by 3 and add the already-present vertical energy to find the total vertical heating 
of the disk.  We assume instantaneous spreading of the energy throughout the star population in the
disk.   When applying Equation 3, we combine $\rho$ and $\lambda$ to give the surface density $\Sigma$ and we consider the following relation for the surface density of the 
stellar disk (Binney \& Merrifield, 1998):

\begin{equation}
\label{equa}
\Sigma (R)\, = \,\Sigma _{0} \,\exp \left( { - \,R\, / R_{d}}  \right)
\end{equation}

\noindent
where R is the radial distance of the impact point, and $R_{d} \, = 
\,2.5$kpc (Binney {\&} Tremaine, 2008).  Combining with the requirement for a total disk mass of  $4.5 \times 10^{10} M_{ \odot}$ (Binney {\&} Tremaine 2008) gives:
\noindent

\begin{equation}
\label{equa}
\rho (R)\, = \,4.46\exp ( - R / 2.5\rm{kpc})$ \textit{M}$_\odot$pc$^{-3}
\end{equation}

 Widrow, Pym \& Dubinski (2008) find independently that $R_{d} \, = 
\,2.8^{+0.23}_{-0.22}$kpc  and  $M_{d} \, = 
\,4.1^{+0.23}_{-0.22} \times 10^{10} M_{ \odot}$.  These values are similar to those used in the potential above.

As the visible disk is about 
15kpc in radius, the calculation discards any crossing beyond that radius 
(there are only just over 200 such cases out of a total number of 10454 
crossings).

To sumarise, the calculation proceeds with the following steps.  First, we calculate the disk's initial vertical kinetic energy based on the mass of 
stars and a velocity dispersion of  $1$kms$^{{\rm -} {\rm 1}}$ (this is $4.5\times10^{53}$ ergs).  We then calculate the energy transfer $E_{T}$ for the first crossing, using Equation 
3, and spread the transferred energy uniformly over all stars in the disk and 
calculate \textit{$\sigma $}$_{z}$, the vertical velocity dispersion based on 1/3 of the 
energy transfer being available for vertical heating.  Finally, we repeat the steps above for all crossings in turn, for all fifty-four GCs.

This leads to a final, present-day \textit{$\sigma $}$_{z}$ = $1.8$kms$^{{\rm -} {\rm 1}}$ from all the crossings by the subset of fifty-four GCs.  Running the calculation with $\omega $Cen only, the most massive GC,  
produces \textit{$\sigma $}$_{z}$ = $1.6$kms$^{{\rm -} {\rm 1}}$, which indicates the importance of GC mass in the energy transfer.  Other factors, related to the GC orbit, influence the energy deposition, mainly the number of crossings, and their location relative to the Galactic centre.  For example, 47Tuc is the second most massive GC in our subset of 54, being a factor 3.3 lighter than $\omega $Cen.  On mass alone, Eqn.3 predicts an energy deposition for 47Tuc, a factor about 11 less than for $\omega $Cen.  In fact, we calculate the energy deposition to be a factor 75 less.  This is because 47Tuc has only 138 crossings, whereas $\omega $Cen experiences 262 crossings.  In addition, the radial position of $\omega $Cen crossings ranges from 1kpc to about 6kpc, with a quarter being near 1kpc.  On the other hand, 47Tuc crossings range from just under 6kpc to just over 7kpc, i.e. in a less dense region.  

To obtain the heating from all 160 GCs, we make the assumption that the 54 orbits and masses are representative of all 160 orbits and masses, and multiply the energy transfer at each crossing in the above calculation by the ratio of 160 to 54.  The result is a present-day disk heating of \textit{$\sigma $}$_{z}$ = $2.8$kms$^{{\rm -} {\rm 1}}$.  This is clearly an overestimate, as it implies that there are $\sim3$ $\omega $Cen equivalents among the current total of 160 GCs.  Nevertheless, this result shows the relatively small effect the current population could have had on disk heating.

We have examined the consequence of regarding the GC as a point mass, whereas in reality it is a system with many degrees of freedom. We 
follow Binney {\&} Tremaine (2008) and find that the total energy 
transferred by disk crossing to the stars in the GC is 

\begin{equation}
\Delta E\, = \,{\frac{{14\pi ^{2}G^{2}\Sigma _{d}^{2} a^{2}M}}{{3v_{M}^{2} 
}}}
\end{equation}where $\Sigma _{d}^{} $ is the surface density of the stellar disk, and $a$ 
the average semi-major axis of stars within the GC. The quantity in square brackets in Eqn.3 is close to unity in all cases considered here.  Hence, the ratio of the 
dynamical friction (Eqn. 3) to the quantity represented by Eqn.10 is then {\Large$ \simeq 
{\frac{{5M}}{{\Sigma_{d} \,a^{2}}}}$}, as 
$\rho \,\,\lambda = \,\Sigma_{d} $. 
With a disk surface density 
of $45M_{ \odot}  \,pc^{ - 2}$, $a$ $\sim$ 1 pc and the 
mass of those GCs that play the overwhelming part in energy transfer $>$ $10^{5} M_{ \odot} $, the ratio of the energy transfer to the disk by dynamical friction compared to the energy transferred to the stars within it is $>$10$^{4}$.  Hence the fraction of the orbital energy diverted into tidal effects is negligible.

\section{Effect of all GCs since Galaxy formation}

While the fraction of the orbital energy diverted to tidal effects is a small fraction of the orbital energy, it is a larger fraction of the binding energy.  Hence the disk transits and other effects in Section 2 contribute to the dissolution of GCs.  The number of GCs in the past was therefore much higher and we now consider the effect of this on GC heating contribution.  We follow all GCs ever present from $\sim$10$^{10}$ years ago, to 
the present, using mass functions from Fall {\&} Zhang (2001).  We select a Schechter function from their figure 3 for the GC-IMF.  This has the form $\psi_0(M) \propto 
      M^{\beta}\exp(-M/M_*)$ with $\beta=-2$ and $M_*=5\times 10^6~M_{\odot}$.  We chose this IMF because it yields a 
close fit to the observed distribution at the current epoch, while including a large component of initially present GCs.  Specifically, using the data in Fall \& Zhang, the  GC-IMF is:

\begin{equation}
\label{eqa}
\psi_0(M)\, \cong 2\times 10^{8}\,M^{ - 2}\,\exp ( - M / M_{ *}  )
\end{equation}and the evolved mass functions are given for t=1.5, 3, 6, and 12 Gyr.  These have a more or less lognormal form, with peaks gradually moving to higher masses.

We again make the assumption that the orbits and crossings of the fifty-four GCs are 
representative of all GCs ever present. It is then possible to calculate the 
disk heating for, say, $N$ GCs with this mass spectrum, by assuming that for each crossing by one of 
the fifty-four, the energy transfer is in fact $N$/54 times larger. 

In the calculations we use 16  bins to cover the mass range $10^3$ to $10^7$$ \textit{M}$$_\odot$.  The total initial number of GCs is $\sim$160,000.  We account for GC mass loss, hence varying mass function, by applying the evolving mass functions of Fall \& Zhang (2001).  As these authors use only four time subdivisions, we interpolate between their curves, for each mass bin, to obtain 14 time subdivisions.
    We calculate the heating due to the complete history of heating caused by these populations depositing energy in the disk through dynamical friction for each energy bin and time interval before summing to obtain the total energy input.  
    
    The above calculations lead to our finding that GCs could on their own heat the disk to a value  \textit{$\sigma $}$_{z}$ = $5.5$kms$^{{\rm -} {\rm 1}}$.  This value lies below the thin and thick disk values of 18 and 39kms$^{{\rm -} {\rm 1}}$ respectively.

These calculations intrinsically carry the time history of the \textit{$ \sigma $}$_{z}$ dispersion.  We have therefore plotted the dispersion as a function of age in Figure 1. Jenkins \& Binney (1990) found that the dispersion increases with age as \textit{$\sigma $}$_{z}$  $\sim \,t^{\alpha}$.  We find from fits to our data that $\alpha$=0.22. We are able to compare this with observational data such as that from Nordstr\"{o}m et al (2004) who performed a kinematic analysis of over 16,000 nearby F and G stars, and who found a value of  $\alpha$$ \sim$ 0.47. 

To obtain a value of \textit{$\sigma $}$_{z}$ = $26$kms$^{{\rm -} {\rm 1}}$, a value predicted for a population aged 10Gyr from the Nordstr\"{o}m data, the energy input from GCs would need be twenty five times larger than provided here by the GCs.  In other words, the GCs are at most a minor contributor to disk heating.  

Our value of \textit{$\sigma $}$_{z}$ can be seen as an upper estimate, because the radial extent of the disk was smaller in the past, and because the number of stars was smaller in the past, two facts unaccounted for here.  On the other hand, orbital decay due to dynamical friction has been neglected, thus also overestimating the energy transfer calculated here, but this neglect has also underestimated the initial GC-IMF.  

We have checked the possible contribution of stars stripped from the GCs by the effects in Section 2 to the vertical velocity dispersion of the disk.  The total mass of stars from the evaporated GCs using Fall \& Zhang's GC-IMF is $ 
\sim 1.5\,\times 10^{9}M_{ \odot}  $, or $<$4$\%$ of the total disk mass.  Because the fraction residing in the disk at any instant is at least an order of magnitude smaller, even if these stars are kinematically hot, their contribution can be neglected.

The particular selection of GC mass functions may have a bearing on the result.  Parmentier \& Gilmore (2007), for example, discuss bell-shaped mass functions.  These are particularly at variance with the GC-IMF we selected from Fall \& Zhang, which rises monotonically as the mass decreases.  To investigate the impact of this, we assume that the GC-IMF we chose, drops precipitously to zero near the peak of the current mass distribution, towards lower values of mass.  We calculate that the energy input would drop only 4\%, with almost no effect on \textit{$\sigma $}$_{z}$.  This underlines the importance of the massive clusters.

\begin{figure}
\includegraphics[bb= 0 360 552.38 1013,width=74mm,clip]{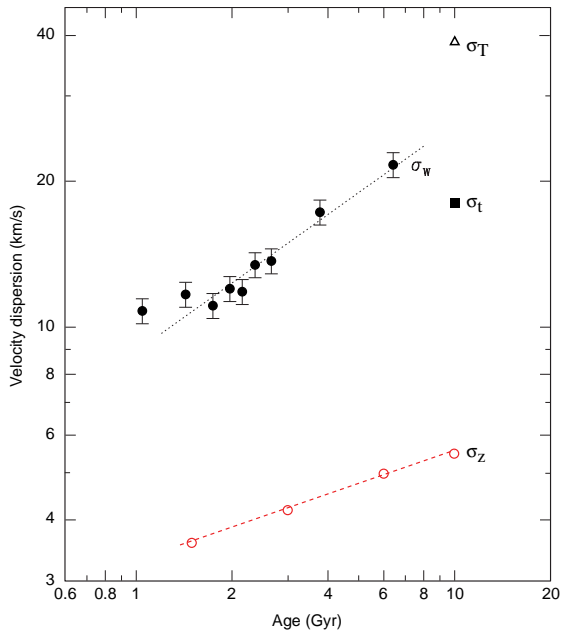}
\caption{Velocity dispersions $ \sigma _{W}$ for single stars with relative age errors $<$ 25\% as a function of age from Nordstr\"{o}m et al (2004, solid circles), with our corresponding values of  $\sigma _{z}$ due to GC impacts.  $\sigma _{T}$ and $\sigma _{t}$ the thick and thin disk values.}
\end{figure}

\section{Conclusion}

We have investigated the vertical heating of the Galactic disk's stars, due to GC crossings inducing dynamical friction.  We achieved this by modeling the orbits of a subset of fifty-four GCs with well-determined orbital parameters, and by assuming a static Galactic potential and disk stellar population.  Assuming these orbits and masses are representative of all GCs at all times, these results were first extrapolated to the 160 existing GCs to find their effect on disk heating.  Next, to consider the effect of all GCs ever present, we used time-dependent mass functions representing all GCs ever present in the Galaxy.  The resulting value of {$\sigma $}$_{z}$ = $5.5$kms$^{{\rm -} {\rm 1}}$ is smaller than that of the thin or thick disks (18 and 39 kms$^{{\rm -} {\rm 1}}$, respectively).  Using the vertical velocity dispersion of Nordstr\"{o}m et al (2004), the GC vertical heating contribution is about one twenty fifth that of the total disk vertical heating, making GCs a second order contributor to Galactic disk heating.  As the energy transfer scales with mass squared, an extension of this latter result has the implication that the upper limit to the total mass associated with a GC is about five times the stellar mass.  This places a constraint on the possible dark matter content in GCs, as far back as the time of Galaxy formation.

\section*{Acknowledgments}

We are grateful to Kinwah Wu and Carlos Allende Prieto for valuable discussions and suggestions.  We are also particularly grateful to the referee for suggesting the result relating to the upper limit on GC mass.  DVP acknowledges the support of an STFC grant.

\end{document}